# Hot-Cold Spots in Italian Macroseismic Data

## G. Molchan[1,2], T. Kronrod[1,2], G.F. Panza[1,3]

*[1] The Abdus Salam International Centre for Theoretical Physics, SAND group, Trieste, Italy*
*[2] International Institute of Earthquake Prediction Theory and Mathematical Geophysics, Russian Academy of Sciences, Moscow*
*[3] Department of Earth Sciences, University of Trieste, Italy*

*Abstract.* The site effect is usually associated with local geological conditions, which increase or decrease the level of shaking compared with standard attenuation relations. We made an attempt to see in the macroseismic data of Italy some other effects, namely, hot/cold spots in the terminology of Olsen (2000), which are related to local fault geometry rather than to soil conditions. We give a list of towns and villages liable to amplify (+) or to reduce (−) the level of shaking in comparison with the nearby settlements. Relief and soil conditions cannot always account for the anomalous sites. Further, there are sites where both (+) and (−) effects are observed depending on the earthquake. The opposite effects can be generated by events from the same seismotectonic zone and along the same direction to the site. Anomalous sites may group themselves into clusters of different scales. All isolated anomalous patterns presented in this paper can be used in hazard analysis, in particular, for the modeling and testing of seismic effects.

*Key words*: site effect, seismic intensity, macroseismic data, seismic hazard

**I. *Introduction.*** The impact of earthquake waves on a structure, referred to as *seismic input* below, is the combined effect of the earthquake rupture process, travel path, and site conditions. The uppermost – unconsolidated layers and the topography make significant contributions to the seismic input. It is a known fact (see, e.g., Mercalli, 1907; Bard & Bouchon, 1980; Shteinberg et al., 1993; Joyner, 2000; Field et al., 2000; Panza et al., 2001; Cornou et al., 2003) that the seismic input is enhanced in shallow valleys owing to the generation of high-amplitude local surface waves. This amplification in deep valleys is due to possible resonances. When the contrast between bedrock and sediment is high, the motion becomes much longer owing to multiple reflections of surface and body waves within the basin (e.g. Zuccolo et al., 2008). When the soil layer is unconsolidated and dipping, seismic rays may experience focusing. This latter effect for a plane seismic wave depends on the angle of emergence, hence is unstable.

Routinely the seismic input is ranked by peak acceleration or velocity in the frequency range 0 to 10 Hz. There are many simple recipes for assessing the seismic input, termed attenuation relations (for a review see Douglas, 2003). Significant deviations – by factors of two or more – of the observed seismic input from the value predicted by these relations are relegated to the category of site effect.

The attenuation relations usually apply to the average soil. For this reason a local departure from the model leads to a regular site effect and is corrected for local soil conditions. However, the origin of site effects is much more diversified, as follows from the above-mentioned facts. Moreover, the anomalous patterns can appear both at different scales and stochastically; one cause of this consists in earth structure complexity, hence the instability of the effect.

Field and SCEC Working Group (2000), Olsen (2000) published 3-D computer simulations of earth ground-motion amplification in southern California. They show that each earthquake exhibits unique "hotspots" of anomalous strong shaking, even when the local geological conditions are taken into account. Hotspots depend on specific details of the earthquake, such as orientation of the fault, irregularities of the rupturing fault surface, and wave scattering, which depends on subsurface structures.

It is natural that a realistic reproduction of this effect requires detailed earthquake and environment information. There are also limitations associated with computational techniques and with local instabilities in the seismic input. In this connection macroseismic data are of independent interest because they contain information on a wide range of seismic input over a large territory. Below we deal with Italian data that are available for a long time span (a few hundred years) and sample almost the entire region.

We remind the essential properties of the macroseismic measure of seismic input, i.e., intensity *I*:
(a) it is not an instrumental quantity; nevertheless, the change in *I* by unity in the MSK- or MCS-type
    scales roughly corresponds to the change in peak acceleration by a factor of 2 to 2.5 (Cancani, 1904;
    Shteinberg et al., 1993; Panza et al., 1997; Shebalin & Aptikaev, 2003); therefore, the deviation $\Delta I$ of

observed *I* from the expected value as large as $|\Delta I| \geq 1$ is here considered anomalous;
(b) *I* is not a point characteristic; it is obviously statistical in character and records the average seismic input to an area of scale *L*=1–10 km (the size of a typical village);
(c) *I* is a discrete quantity on an integer or half-integer scale. Non integer values are often reported to indicate the difficulty to assign the observed effect to a specific intensity degree.
The properties (b) and (c) grant certain stability to *I*.

In the regular situation macroseismic intensity decays with distance, therefore, isoseismal areas of level *I* (the area with intensities of at least *I*) are embedded in isoseismals area of level *I*-1. A local anomaly of macroseismic intensity can disturbs the simple connectivity of an isoseismal area and makes *I* to deviate from the expected value, i.e., produces *inversion* of *I* in the terminology of Shebalin (2003). The inversion of *I* may be positive or negative, isolated or clustered. We plan to analyze these features in Italian macroseismic data because they have to include the "hot/cold spots" patterns by Field & SCEC Working Group (2000).

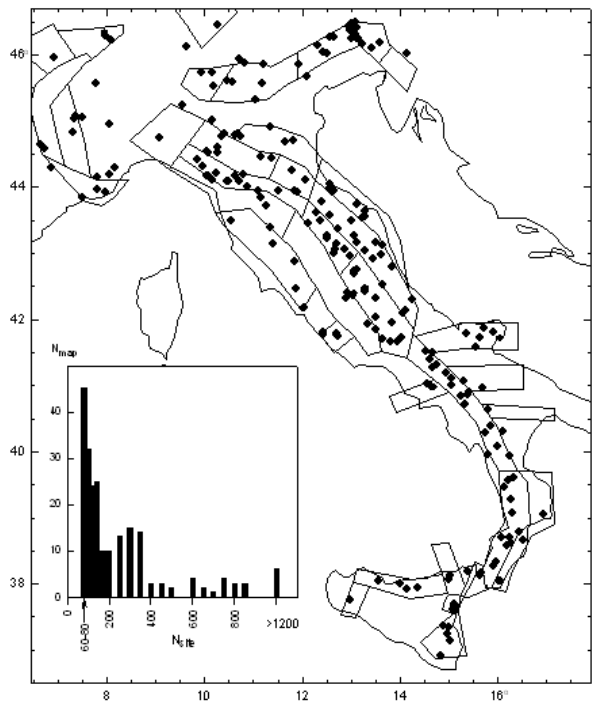

Fig. 1. Earthquakes selected for the analysis of *I*-maps.
The inset shows the number of maps ($N_{map}$) with given number of sites ($N_{site}$). The *thin lines* show the seismic source zones by Miletti et al. (2008).

## 2. The Data.
The CFTI catalog (Boschi et al., 2000) is used as the main source of Italian macroseismic data. This has been supplemented by data taken from the DOM database (Monachesi and Stucchi, 1997), Bolletino Macrosismico (BM, 1988-1993), and from the new DBMI04 database (2004).

For the purposes of this work we selected 229 *I*-maps, $I(g)$, using the following criteria:
• the number of observation sites in a map should be $N_1 \geq 60$;
• the map should contain observations of at least three integer-value levels;
• the observation sites should cover an area of linear extent $L > 100$ km.
As a result a typical *I*-map contains from 60 to 300 sites. For most of the selected events isoseismal maps have been produced in our previous publications (Kronrod et al., 1999, 2002).

To relate the data to the seismotectonic zones we use the Italian seismic source zone model by Meletti et al. (2008).

Figure 1 shows that the epicenters of the selected events provide a fairly good coverage of the seismogenic areas of Italy.

## 3. Empirical Analysis.
With the exclusion of the regular site effect, Olsen (2000) in his 3-D computer simulations reproduce anomalous strong shaking patterns with the following properties: they are non-regular, isolated, and are ≤30 km in extent. Similarly, we would like to find such effects in macroseismic data. With this aim, primarily we identify the sites where isolated local *I*-inversions have been observed in the past. Next, for each selected site we classify the *I*-maps that contain this site.

*First step.* Choosing an $I(g)$ map, we classify a site $g_0$ on $I(g)$ as anomalous if

1) $g_0$ is outside the first isoseismal;

2) $I(g_0) \geq \mathrm{IV}$;

3) $g_0$ has a sufficiently dense surrounding of nearby sites $g_k$ relative to which $I(g_0)$ is an isolated local anomaly, i.e., as a rule, the deviation of $I$ is large and has a uniform sign.

More exactly, the last requirement means that:

3a) the number of nearby sites $\{g_k\}$ is at least $N_-$, i.e., $N(g_0) \geq N_-$; in the following we take $N_- = 5$;

3b) the nearby sites are enclosed within a ring: $2 < |g_0 - g_k| < R$, where $R \leq 40$ km;

(The choice $R \leq 30$ km, given $N_- = 5$, has the result that the anomalous sites will be rare in Italy.)

3c) all angles, $\alpha_k$, between adjacent vectors $\boldsymbol{g_0 g_k}$ are less than $\alpha_-$.

To choice of $\alpha_-$, note that

$$360° = \Sigma\alpha_k \leq \alpha_- \cdot N(g_0),$$

therefore we can determine $\alpha_- = 72°$ because $N(g_0) \geq 5$;

3d) at least $N(g_0) - 2$ nearby sites have identical values of

$$s(g) = \mathrm{sign}\ (I(g_0) - I(g)),$$

say $s(g_k) = s$. All sites with $s(g) \neq s$ are considered to be "*noise*".

We admit the equality $I(g_k) = I(g_0)$ at one site at the most and the inequality

$$|I(g_0) - I(g_k)| < 1. \tag{1}$$

for 10% of the sites, at the most.

Thus, a $g_0$ site on an $I$-map acquires the sign $s(g_0|I) = s$ with values (+) or (−). If all requirements (1–3) are fulfilled with the exception of (3d) for any $R \leq 40$ km, then to the quantity $s(g_0|I)$ the value 0 (normal) is assigned. In the other cases $s(g_0|I)$ is treated as indeterminate.

Finally, the isolated site effect on an $I$-map is characterized by an anomalous deviation of $I(g_0)$:

$$s(g_0|I)\ \min'_k\ |I(g_0) - I(g_k)| = \Delta I(g_0). \tag{2}$$

Here the minimum is taken over all sites $g_k$ with the exception of the "noise" sites in (d) and the sites which satisfy (1).

Examples of (+) and (−) sites are given in Fig. 2.

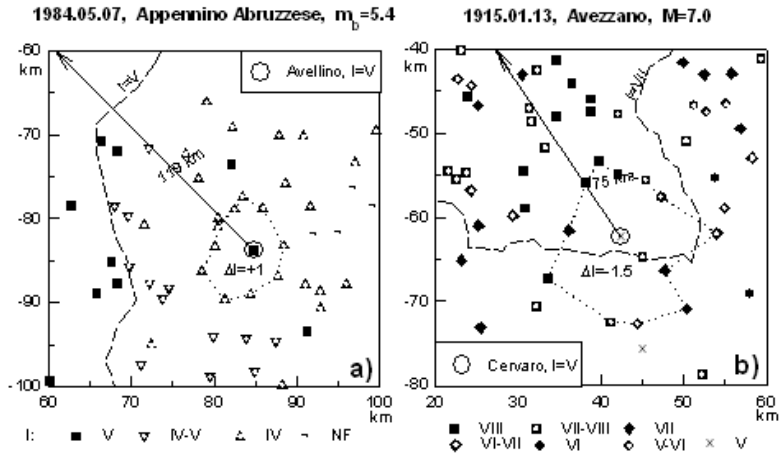

Fig. 2. Examples of isolated local anomalous sites:

(a) hot, (b) cold.

The *dashed line* indicates an isoseismal of level $I$; the *dotted line* delimits the neighborhood of an anomalous site. The epicenter is used as the zero point in the Cartesian coordinates (in km). Notation 75↓ means direction and distance in km to an epicenter.

*Second step.* For each selected anomalous site $g_0$ we revise all $I$-maps again and choose the maps that satisfy conditions 1 and 3 at $g_0$. In other words, the requirement $I(g_0) \geq 4$ is omitted to increase the number of maps covering $g_0$. As a result we have at $g_0$

$$\max_I I(g_0) \geq \mathrm{IV} \geq \min_I I(g_0), \tag{3}$$

where the extremes are taken from the maps with $s(g_0|I) \neq 0$.

The subsequent discussion is limited to the sites for which

4) $N_{\mathrm{map}}(g_0) \geq 7$, where $N_{\mathrm{map}}$ is the number of maps for which $s(g_0|I)$ have been assigned.

Now each anomalous site $g_0$ can be characterized by the following quantity:

$$f(g_0) = N_{\mathrm{map}}(g_0)\ /\ N_{\mathrm{a}}(g_0), \tag{4}$$

where $N_{\mathrm{a}}(g_0)$ is the number of maps with $s(g_0|I) \neq 0$. The quantity $f(g_0)$ means that the local $I$-effect at $g_0$ is observed on average in each $f(g_0)$-th map for which $s(g_0|I)$ have been assigned.

The anomalous sites are divided into three categories:

♦ those with a persistent nonnegative effect ("hot", +):

$$s(g_0|I) \geq 0 \text{ for any } I\text{-map} \tag{5}$$

♦ those with a persistent nonpositive effect ("cold", −):

$$s(g_0|I) \leq 0 \text{ for any } I\text{-map}; \tag{6}$$

♦ those with a mixed effect (hot/cold or +/−), i.e. there exist both maps with $s(g_0)$=+ and maps with $s(g_0)$=−.

Histograms of $f(g)$ for the three categories of sites are given in Fig. 3. They show that the typical range of $f(g)$ is (4, 10), i.e., typically the hot or cold effect is not a permanent feature of the anomalous site. The same property holds for the "hotspot" effect by Field and SCEC Working Group (2000). But our definition of an isolated local anomalous site is based on local features in the set of $I$-maps that cover the site, and thus is independent of any attenuation relation.

The three categories contain very different numbers of anomalous sites:

$$\#(+) = 65; \quad \#(-) = 21; \quad \#(+/-) = 10. \tag{7}$$

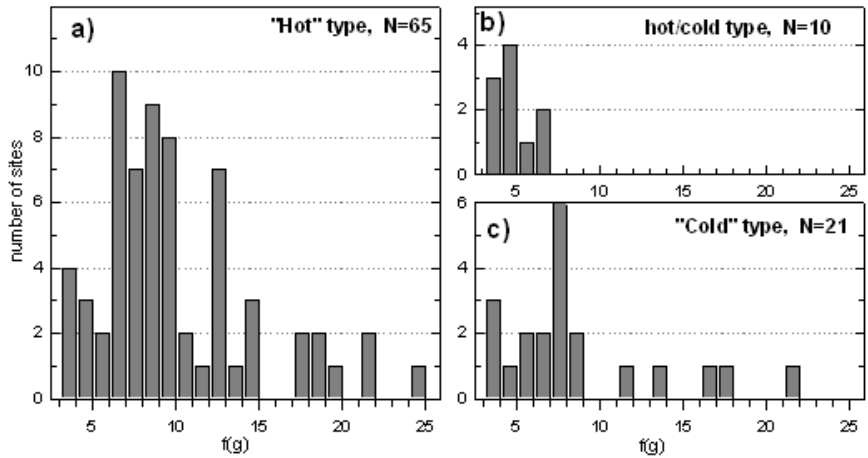

Fig. 3. Histograms of $f(g) = N_{map}/N_a$ for three types of sites: (a) hot (+), (b) mixed (+/−), and (c) cold (−).

The cause of this is twofold; on the one hand the type (−) sites were of less interest at the time the macroseismic effects were recorded, hence are less complete; on the other hand, our analysis of isolated local anomalies is asymmetrical with respect to the (+) and (−) types. The first step of analysis is concerned with the sites having $I(g_0) \geq$ IV. When $I(g_0)$=IV, a vicinity of $g_0$ must contain sites with $I(g) \leq$III if it is a (+) effect and with $I(g) \geq$ V if it is a (−) effect. These requirements are not equivalent and restrict the set of (−) sites.

Figure 4 and Table 1 summarize the anomalous sites that have been identified. They provide the spatial distribution of the sites, the characteristic $f(g_0)$, all events that have given rise to the anomaly at a site, the magnitude of the anomaly $\Delta I(g_0)$, and the epicenter-site distance. It is possible to conclude that an isolated anomalous site in Table 1 has the following typical properties:

– it is located 50-150 km from the epicenter;

– the radius $R$ of the neighborhood with respect to which $g_0$ is anomalous is 10-40 km (the upper bound, 40 km, is a consequence of condition 3b);

– the number of sites around $g_0$ is 10–20;

– the number of $I(g)$ maps that cover $g_0$ is 7–20 (the lower bound, 7, is a consequence of condition 4).

Overall, the (+) and (−) sites cover the region uniformly enough, with the exception of the western Alps where the density of observation is low.

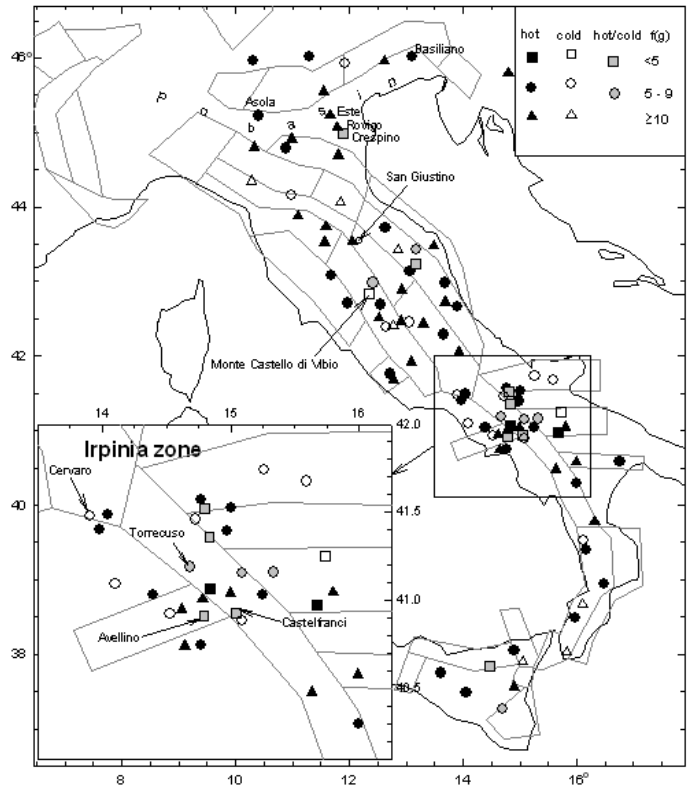

Fig. 4. Anomalous sites of three types (hot, cold, hot/cold) and their characteristic *f(g)*; the *thin lines* show the seismic source zones by Meletti et al. (2008).

## 4. *Individual examples*

*4.1. Isolated local anomalies .*In most cases the explanation of local anomalous effects is not obvious and not unique. For example, hot and cold effects would be naturally expected where an appreciable soil contrast exists between the site itself and its environs. A few sites in Fig. 4 fulfill these expectations. One of the more positive examples is the Monte Castello di Vibio site of type (−) (see Fig. 4). This site is on hard rock, while sediments cover the environs. The relationship is just the reverse for Avellino and its environs (Fig. 4). For this reason the hot effect is typical there (see Fig. 5a). However, the Gargano earthquake of August 12, 1889 (N 1 on Fig. 5a) produced *I* = II-III at Avellino and *I*=IV &_*I*=IV-V for the environs, i.e., the cold effect, contrary to expectation. In principle, a case of this type is possible due to the influence of previous earthquakes on the vulnerability of structures (Baratta, 1906). But it is unlikely for the isolated effect.

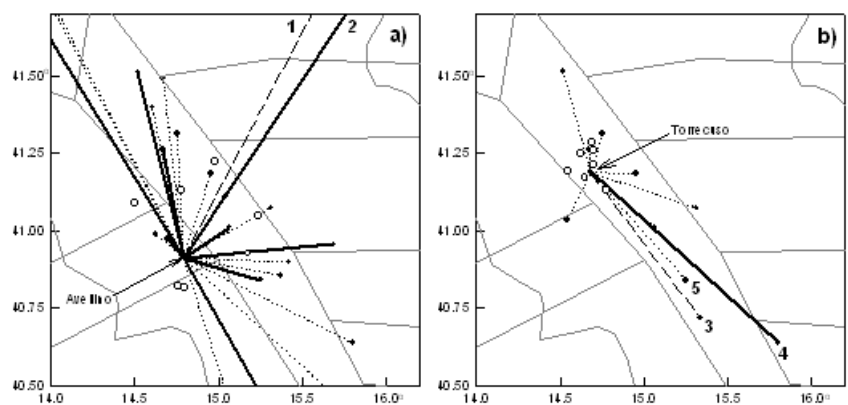

Fig. 5. Sites with (+/−) type effect: (a) Avellino and (b) Torrecuso.
The lines (*bold*, *dotted* or *dashed*) connect the sites with events and indicate the type of local effect ((+), 0 or (−) respectively); *open circles* represent *I*-data points from the site neighborhood, which are encountered in all *I*-maps where the site is anomalous. Additional information on the events with numbers assigned in the plot:
1) 1889.12.08, Apricena, dist=120 km, *I*(g$_0$)=2.5, the local effect Δ*I*=−1;
2) 1951.01.16, Gargano, dist=125 km, *I*(g$_0$)=3.5, Δ*I*=+0.5 (doubtful (+)-effect);
3) 1990.05.05, Potenza, dist=83 km, *I*(g$_0$)=no felt, Δ*I*=−3;
4) 1991.06.26, Potentino, dist=121 km, *I*(g$_0$)=5, Δ*I*=+1;
5) 1980.11.23, Irpinia-Lucania, dist=53 km, *I*(g$_0$)=6, Δ*I*=0.

The sites of (+/–) type are the most interesting for interpretation because the traditional explanation in terms of soil conditions and the thickness of sediment above hard bedrock is insufficient. Fig. 5b (see also Fig. 4) represents the Torrecuso site. Here three events have produced different effects at the site, namely (+), (–), and "normal" (0); at the same time they have similar directions toward the site.

In the deep sediment-filled Po Basin (see Fig. 4), the expected effect relative to the normal soil ought to be of type (+). This is confirmed by many events (see at least ten *I*-maps in Kronrod et al., 1999) occurring in the surrounding area because their isoseismals anomalously extend into the sediment zone. We are however interested in the isolated local *I*-effects observed in the Po Basin (see Fig. 4). In this particular case the site and its environs are in "identical" soil conditions, hence the type of the anomalous site is not clear beforehand. Fig. 4 shows four sites of type (+) and one site of type (+/–) in the Po Basin, i.e., isolated local anomalous sites are available here.

*2. Clustered anomalous effects*. Intensity inversion can also occur at large scales, some tens or hundreds of kilometers. In that case we should call the phenomenon a *collective or clustered anomalous effect*. Such effects can not be easily formalized. Therefore we will consider a few examples, which may be of interest for the simulation and testing of Earth models.

Fig. 6 represents two *I*-maps with a collective anomalous effect in the Po Basin. In the case of the 1898, Calestano, M=4.7 event (Fig. 6a), the isoseismal of level *I*=III is overextended into the sediment zone (domain **A**). To interpret this pattern, we use a recent paper by Carletti and Gasperini (2003), who suggest a nonhomogeneous *I*-attenuation model for the entire territory of Italy. Namely, the increment of *I* between an epicenter ($g_0$) and a receiver (site $g$) is given by the following empirical relation

$$I(g_0)\text{-}I(g)=\varphi(|g_0\text{-}g|) \quad +\int_0^{s_0} b_1(g_0 + s(g-g_0))ds + \int_{s_0}^1 b_2(g_0 + s(g-g_0))ds \qquad (8)$$

where $s_0=\min(1, d/|g_0\text{-}g|)$ and $d$=45 km; the first part of (8), $\varphi(|\Delta g|)$, is an average isotropic attenuation relation for Italy while $b_i(g)$, $i$=1, 2 are responsible for the spatial variations of the relation in near (the case $b_1$) and far (the case $b_2$) zones relative to $g_0$. In particular, a point $g$ with $b_2(g)$<0 is a point of low attenuation of *I* with respect to the average model in the zone: $|g\text{-}g_0|>d$. The functions $b_1$ and $b_2$ are computed on meshes of 50 and 25 km.

Fig. 6 shows two isolines of $b_2(g)$ in the subarea of the Po Basin where $b_2(g)$<0 ($|b_2(g)|$ increases toward northeast). In the case of the Calestano event (Fig. 6a) the model (8) well explains the anomalous part of the *I*=III zone (domain **A**), namely, the shape and level of that zone.

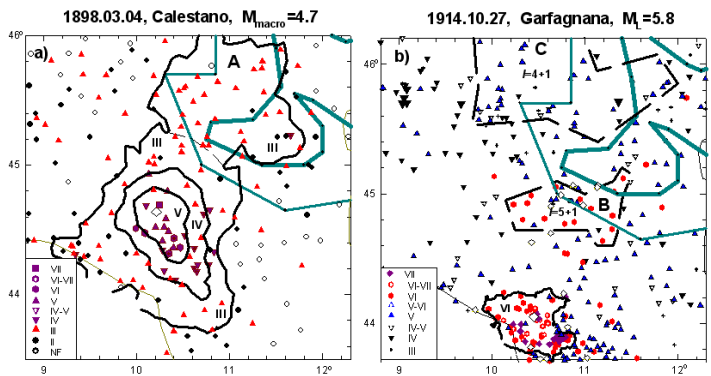

Fig. 6. Examples of clustered anomalous patterns in the Po Basin for the following events: (a) 1898, Calestano, M=4.7, and (b) 1914, Garfagnana, M=5.8.
Isoseismals (*bold line*) are based on the Modified Polynomial Filtering (MPF) method by Molchan et al. (2004); domain **A** is an anomalous part of the *I*=III zone for the Calestano event, a natural boundary of the *I*=III zone is given by a *dashed thin line*; domains **B** and **C** are zones of collective *I*-inversion for the Garfagnana event. The *open diamond* marks the epicenter.
*Gray line* is the level of low-attenuation of *I* according to Carletti & Gasperini (2003) (see main text).

The *I*-map for the 1914, Garfagnana earthquake, *M*=5.8 (Fig. 6b) has anomalous features practically in the same place. The isoseismal zone of level *I*=VI consists of two parts, the main and the isolated anomalous zone **B** of type (+). There is also zone **C**, which looks as an anomalous isolated *I*=IV zone of the same type (+). Therefore the pattern **B**∪**C** can be interpreted as a result of the expansion of the highest seismic intensities into the sediment zone of the Po Basin. As Fig. 6b shows, the model (8) is in agreement with the pattern **B**∪**C** but only qualitatively. This model can not explain both the disconnection of the isoseismal zone {*I*≥VI} and the *I*-level of the pattern **C**

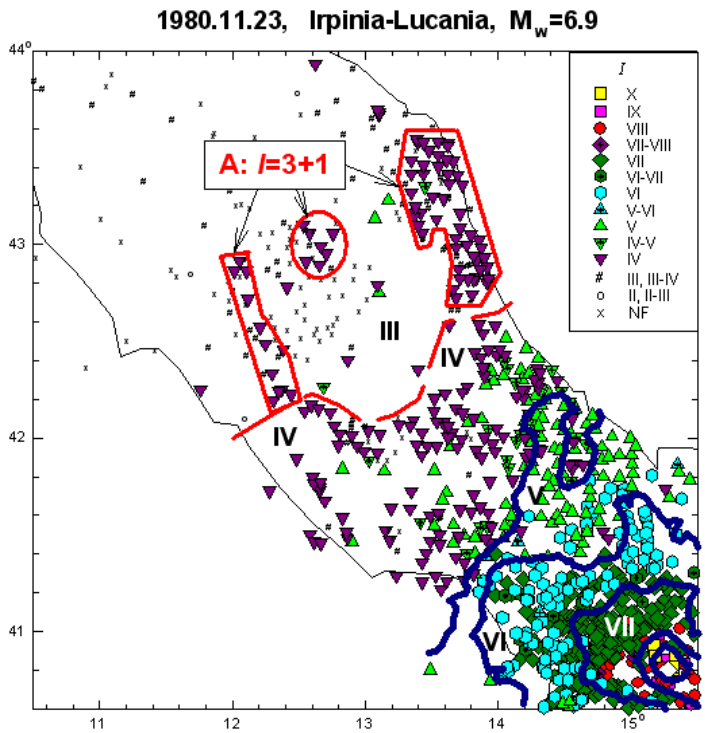

**1980.11.23, Irpinia-Lucania, M_w=6.9**

Fig.7. *I*-map (northern part) for the 1980, Irpinia-Lucania, *M*=6.9 earthquake. Isoseismals (*bluy lines*) are based on the MPF method by Molchan et al. (2004). Domain **A** (*bold red lines*) is an anomalous part of *I*=IV zone; a natural boundary of *I*=IV zone is given by *dashed red line*. The *open diamond* marks the epicenter.

An interesting example with an overextended *I* zone is provided by the *I*-map of 1980, Irpinia-Lucania, *M*=6.9 event (Fig.7). The anomalous part of the *I*=IV zone consists of three disconnected pieces (see zone **A**) and lies in the far zone relative to the epicenter. The model (8) is non-informative at all in this case. The pattern **A** is unique for the Umbria-Marche domain (see the Atlas by Kronrod et al., 2002) and therefore it is interesting for interpretation.

The next two examples (Fig. 8, Fig. 9) illustrate an isolated collective anomalous effect of size 30-50 km. The first one is the *I*-map of the 1987, Porto San Giorgio earthquake (Fig. 8). Here, the {*I*≥V}-zone has a well-pronounced anomalous part **B** of size *L*=45 km; **B** is within the *I*=III isoseismal. The anomaly is situated in the northern Umbria-Marche Apennines and is adjacent to the transition zone that separates the central and the northern Apennines. The sites showing the (+) effect (Δ*I* =2) are situated at different altitudes, between 270 m and 430 m (ref. GG), and on different soils, ranging from limestone and shale to Quaternary alluvium. Model (8) is again unable to explain the isolated effect. In addition, it cannot explain the strong effect (Δ*I*=+2) in **B**.

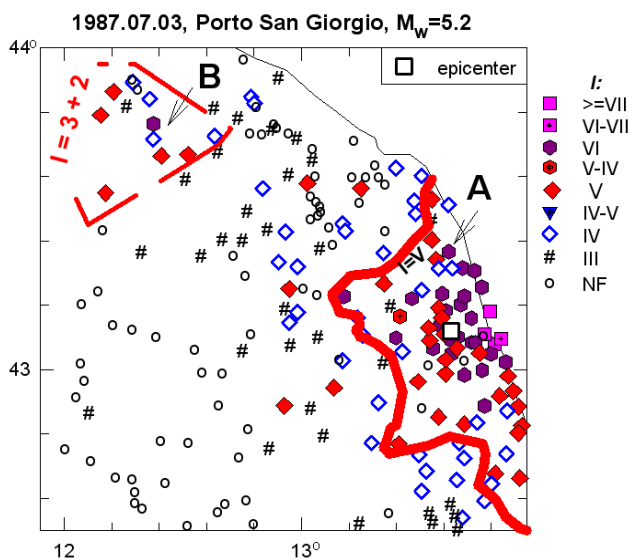

**1987.07.03, Porto San Giorgio, M_w=5.2**

Fig. 8. Example of isolated anomalous patterns of type (+): *I*-map of 1987, Porto San Georgio, *M*=5.2 earthquake. The {*I*≥V} zone consists of two parts: main (**A**) and anomalous pattern (**B**); the effect in **B** is Δ*I*=+2.

The *I*-map of the 1984, Apennino Abruzzese, *M*=5.9, earthquake is another example of the isolated collective anomalous effect but now of type (−) (see Fig.9). This effect is localized in a wedge-like zone between the W-E Circeo-Vulture line and the NNW-SSE Appenine fault system. The sites within **C** are situated at different altitudes, between 297 m and 981 m (ref. GG), on slopes with different steepness, and in different soil conditions ranging from volcanic bedrock to young sedimentary deposits (CNR P.F. Geodinamica, 1992).

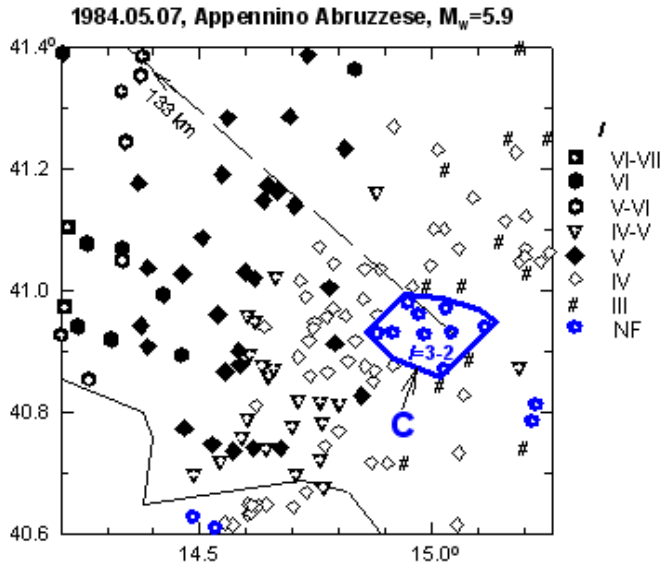

Fig. 9. Example of isolated collective anomaly of type (−): *I*-map of the 1984, Apennino Abruzzese, *M*=5.9 earthquake; **C** is anomalous zone of type (−); the effect in **C** is Δ*I*=−2; ↑133 km indicates direction and distance from zone **C** to the epicenter.

To understand the effect in **C** we considered twenty *I*-maps that cover **C**. The epicenters of the events concerned are plotted in Fig. 10. All events to the north of the Circeo-Vulture fault system have only the non-positive effect in **C**. To be specific, 3 events out of 8 produce normal effect and 5 produce a cold spot. The other 12 events produce normal effect in **C** and occur either on the Circeo-Vulture faults or to the south of **C**. Thus, we may suppose that the cause of the (−) effect in **C** is the strong attenuation of the seismic waves travelling across the Circeo-Vulture fault system from north to south. In the framework of model (8) we should expect the (+) effect at those sites of **C** where (−) has been observed. Indeed, let us connect each event which generates a (−) effect in **C** with **C** by a segment (see Fig. 10). On the figure we can see the trace of the intersection of the segment with the low *I*-attenuation zone according to (8); since the trace dominates in each segment, we have a (+) effect in **C**.

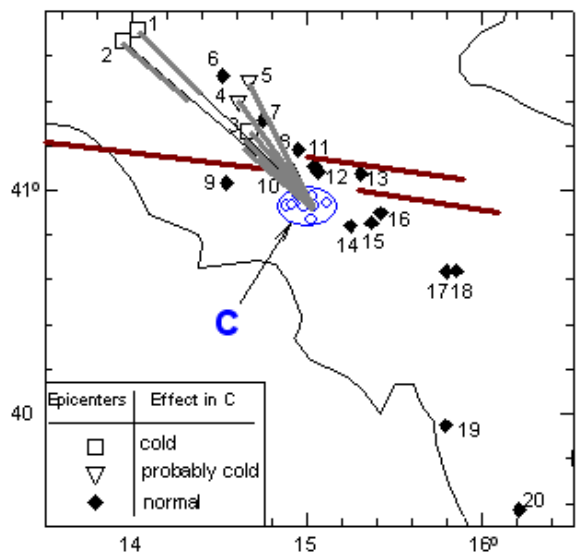

Fig. 10. The anomalous zone **C** (oval) shown in Fig. 9 and 20 events whose *I*-maps cover **C**:
1) 1984.05.07, M=5.9;  2) 1984.05.11, M=5.5;  3)1688.06.05, M=6.6;  4) 1456.12.05, M=7.1;  5) 1913.10.04, M=5.2;  6) 1805.07.26, M=6.7;  7) 1997.03.19, M=4.6;  8) 1962.08.21, M=6.2;  9) 1903.05.04, M=5.0;  10) 1905.03.14, M=4.7;  11) 1905.11.26, M=5.1;  12) 1732.11.29, M=6.4;  13) 1930.07.23, M=6.7;  14) 1980.11.23, M=6.9;  15) 1694.09.08, M=7.0; 16) 1910.06.07, M=5.9;  17) 1991.05.26, M=4.7;  18) 1990.05.05, M=5.8;  19) 1982.03.21, M=5.0; 20) 1887.12.03, M=6.4.

**C** consists of sites (*blue open circles*). Each event, which generates a (−) effect in **C**, is connected with **C** by segment.

A trace of the low *I*-attenuation zone by Carletti & Gasperini (2003) on the segment is marked by *gray color*.

*Bold brown lines* show the Circeo-Vulture fault system

**5. Conclusion.** The *site effect* is usually associated with local geological conditions, which increase or decrease the level of shaking in comparison with standard attenuation relations. We made an attempt to evidence in macroseismic data some other effects namely *hot/cold spots* in the terminology of Olsen (2000), which is related to local fault geometry rather than to soil conditions. Our analogue of this effect (*isolated I-inversion*) is based on local features in the set of *I*-maps that cover a site, and thus is independent of the reference attenuation model.

The local *I*-inversions are not persistent. They are not very frequent (roughly in 10-25% of the considered cases where abnormal intensities are identified); the relative effect ($\Delta I$) may be either +1 or −1 for different earthquakes. It thus appears that the topography and soil conditions are necessary, but far from sufficient for an interpretation of the *I*–inversions. Thus, our collection of *I*-inversions (Table 1) includes the hot/cold effects as defined by Olsen (2000).

Engineering studies of site effects are relevant to the near zone and $I \geq$ VIII (Shteinberg et al., 1993). Our examples exclude the first isoseismal (the near zone) and are limited to the range $I<$X (MCS) and distances not exceeding 200 km, consequently they serve as a complement to engineering experience.

Intensity inversion can occur at large scales, some tens or hundreds of kilometres as well. The non-homogeneous attenuation model by Carletti & Gasperini (2003) could be useful in the interpretation of such patterns, however, the examples show that this model is far from being always applicable. Therefore, all our *I*-inversion patterns (see Table 1, Figures 5–9) are of interest for the simulation and testing of the seismic input. The list of anomalous sites given in Table 1 is incomplete because of the stringent selection criteria used and the classification of the sites as hot, cold or hot/cold may be modified as more data become available.

*Acknowledgements.* We are grateful to the anonymous reviewers for the careful reading of the paper and very useful comments and suggestions. This work was supported by the Russian Foundation for Basic Research through grant 08-05-00215, and by the ASI-Pilot project SISMA

## Table 1. Anomalous sites and their characteristics

| # | Site $g_0$ (province) | coordinates | | type | $f(g)=$ | $I$-map[*] | | | | |
|---|---|---|---|---|---|---|---|---|---|---|
| | | Lat | Long | $s(g_0)$ | $N_{map}/N_a$ | EQ date | $I_0$ | dist | $I(g_0)$ | $\Delta I(g_0)$ |
| 1 | Accadia (FG) | 41.158 | 15.334 | +/– | 14/2 | 1990.05.05 | 7 | 49 | 4.5 | +1 |
| | | | | | | 1984.05.07 | 8 | 137 | 5 | –1 |
| 2 | Acquasparta (TR) | 42.690 | 12.546 | + | 10/2 | 1984.05.11 | 7 | 175 | 4 | +1 |
| | | | | | | 1984.04.29 | 7 | 65 | 5 | +1 |
| 3 | Altavilla Irpina (AV) | 41.006 | 14.779 | + | 10/1 | 1984.05.07 | 8 | 110 | 5 | +1 |
| 4 | Anghiari (AR) | 43.540 | 12.054 | + | 13/1 | 1920.09.07 | 9.5 | 173 | 5 | +1 |
| 5 | Ariano Irpino (AV) | 41.153 | 15.089 | + | 18/1 | 1456.12.05 | 11 | 52 | 9.5 | +1 |
| 6 | Asola (MN) | 45.221 | 10.413 | + | 10/2 | 1920.09.07 | 9.5 | 112 | 5.5 | +1 |
| | | | | | | 1987.06.02 | 6 | 55 | 5 | +1 |
| 7 | Asolo (TV) | 45.801 | 14.791 | + | 12/1 | 1987.05.02 | 6 | 137 | 4 | +1 |
| 8 | Avellino (AV) | 40.914 | 14.791 | +/– | 24/6 | 1732.11.29 | 9.5 | 24 | 9 | +1 |
| | | | | | | 1805.07.26 | 10 | 74 | 8 | +1 |
| | | | | | | 1889.12.08 | 7 | 120 | 2.5 | –1 |
| | | | | | | 1980.11.23 | 9.5 | 40 | 8 | +1 |
| | | | | | | 1982.03.21 | 7.5 | 149 | 5 | +1 |
| | | | | | | 1984.05.07 | 8 | 119 | 5 | +1 |
| 9 | Bagaladi (RC) | 38.026 | 15.821 | – | 12/1 | 1783.03.28 | 10 | 93 | 5 | –1 |
| 10 | Baiano (AV) | 40.951 | 14.618 | + | 11/1 | 1910.06.07 | 9 | 68 | 7 | +1.5 |
| 11 | Baronissi (SA) | 40.746 | 14.770 | + | 9/1 | 1910.06.07 | 9 | 54 | 7 | +1 |
| 12 | Baselice (BN) | 41.393 | 14.973 | + | 9/1 | 1991.05.26 | 7 | 118 | 4.5 | +1 |
| 13 | Basiliano (UD) | 46.013 | 13.109 | + | 8/1 | 1988.02.01 | 6 | 27 | 5.5 | +1 |
| 14 | Breno (BS) | 45.956 | 10.303 | + | 9/1 | 1936.10.18 | 9 | 161 | 6 | +1 |
| 15 | Brienza (PZ) | 40.478 | 15.628 | + | 10/1 | 1905.09.08 | 11 | 210 | 6 | +1 |
| 16 | Caltanissetta (GL) | 37.490 | 14.057 | + | 8/1 | 1990.12.13 | 7.5 | 89 | 5.5 | +1 |
| 17 | Camerino (MC) | 43.135 | 13.068 | + | 28/3 | 1980.11.23 | 9.5 | 338 | 5 | +1 |
| | | | | | | 1984.05.07 | 8 | 190 | 4 | +1 |
| | | | | | | 1984.05.11 | 7 | 189 | 4 | +1 |
| 18 | Campli (TE) | 42.726 | 13.686 | + | 13/1 | 1984.05.07 | 8 | 124 | 5 | +1 |
| 19 | Cantalice (RI) | 42.466 | 12.904 | + | 13/1 | 1987.07.03 | 7 | 87 | 4 | +1 |
| 20 | Capizzi (ME) | 37.848 | 14.479 | +/– | 9/2 | 1905.09.08 | 11 | 157 | 6 | +1 |
| | | | | | | 1908.12.28 | 11 | 104 | 4 | –1 |
| 21 | Carpi (MO) | 44.784 | 10.885 | + | 17/2 | 1919.06.29 | 9 | 111 | 6 | +2 |
| | | | | | | 1920.09.07 | 9.5 | 77 | 6 | +1 |
| 22 | Cassano allo Ionio (CS) | 39.784 | 16.317 | + | 10/1 | 1980.11.23 | 9.5 | 162 | 6 | +1 |

| # | Site $g_0$ (province) | coordinates | | type | $f(g)=$ | $I$-map[*)] | | | | |
|---|---|---|---|---|---|---|---|---|---|---|
| | | Lat | Long | $s(g_0)$ | $N_{map}/N_a$ | EQ date | $I_0$ | dist | $I(g_0)$ | $\Delta I(g_0)$ |
| 23 | Castel Giorgio  (TR) | 42.708 | 11.979 | + | 7/1 | 1993.06.05 | 6 | 69 | 5 | +1.5 |
| 24 | Castelfranci (AV) | 40.931 | 15.043 | +/– | 8/2 | 1990.05.05 | 7.0 | 37 | 6.5 | +1.5 |
| | | | | | | 1991.05.26 | 6.5 | 76 | 1 | –2 |
| 25 | Cercemaggiore (CB) | 41.460 | 14.722 | – | 9/1 | 1688.06.05 | 11 | 21 | 6 | –1 |
| 26 | Cervaro (FR) | 41.481 | 13.904 | – | 8/1 | 1915.01.13 | 11 | 75 | 5 | –1.5 |
| 27 | Collagna (RE) | 44.347 | 10.275 | – | 18/1 | 1920.09.07 | 9.5 | 15 | 7 | –1 |
| 28 | Colle Sannita (BN) | 41.361 | 14.833 | + | 13/2 | 1984.05.07 | 8 | 86 | 6 | +1 |
| | | | | | | 1984.05.11 | 7 | 83 | 6 | +1 |
| 29 | Concordia sulla Secchia (MO) | 44.914 | 10.981 | + | 10/2 | 1891.06.07 | 9 | 72 | 6 | +1 |
| | | | | | | 1971.07.15 | 7.5 | 46 | 7 | +1 |
| 30 | Contigliano (RI) | 42.411 | 12.769 | – | 14/1 | 1979.09.19 | 8 | 39 | 4.5 | –1 |
| 31 | Cossignano (AP) | 42.983 | 13.688 | + | 7/1 | 1915.01.13 | 11 | 104 | 7.5 | +1 |
| 32 | Crespino (RO) | 44.982 | 11.885 | + | 7/1 | 1987.05.02 | 6 | 93 | 5 | +1 |
| 33 | Deruta (PG) | 42.982 | 12.420 | +/– | 14/2 | 1930.10.30 | 8.5 | 93 | 5 | +1 |
| | | | | | | 1998.03.26 | 6.5 | 57 | 4 | –1 |
| 34 | Este (PD) | 45.228 | 11.656 | + | 15/1 | 1914.10.27 | 7 | 134 | 6 | +1 |
| 35 | Ginosa (TA) | 40.578 | 16.758 | + | 7/1 | 1905.09.08 | 11 | 209 | 6 | +1 |
| 36 | Grazzanise (CE) | 41.091 | 14.102 | – | 8/1 | 1984.05.07 | 8 | 68 | 5 | –1 |
| 37 | Jelsi (CB) | 41.518 | 14.796 | +/– | 8/2 | 1915.01.13 | 11 | 129 | 6.5 | +1.5 |
| | | | | | | 1980.11.23 | 9.5 | 90 | 4 | –1 |
| 38 | Lavello  (PZ) | 41.046 | 15.795 | + | 10/1 | 1857.12.16 | 10.5 | 75 | 4 | +1 |
| 39 | Lercara Friddi  (PA) | 37.748 | 13.603 | + | 7/1 | 1908.12.28 | 11 | 180 | 7 | +1 |
| 40 | Levico Terme  (TN) | 46.011 | 11.303 | + | 8/1 | 1976.12.13 | 7 | 38 | 6 | +1 |
| 41 | Marcianise  (CE) | 41.033 | 14.395 | + | 7/1 | 1915.01.13 | 11 | 139 | 7 | +1 |
| 42 | Micigliano (RI) | 42.451 | 13.054 | – | 9/1 | 1915.01.13 | 11 | 63 | 5 | –1 |
| 43 | Miglierina (CZ) | 38.947 | 16.471 | + | 7/1 | 1990.05.05 | 7 | 237 | 4 | +1 |
| 44 | Mignano Monte Lungo (CE) | 41.404 | 13.983 | + | 8/1 | 1980.11.23 | 9.5 | 132 | 7 | +1 |
| 45 | Mineo  (CT) | 37.266 | 14.691 | +/– | 10/2 | 1905.09.08 | 11 | 186 | 5 | +3 |
| | | | | | | 1980.01.23 | 5.5 | 43 | 2 | –1 |
| 46 | Mirabella Elcano (AV) | 41.042 | 14.996 | + | 13/1 | 1980.11.23 | 9.5 | 33 | 8 | +1 |
| 47 | Moio Alcantara  (ME) | 37.899 | 15.051 | – | 11/3 | 1905.09.08 | 11 | 117 | 3 | –1 |
| | | | | | | 1908.12.28 | 11 | 57 | 5 | –1 |
| | | | | | | 1975.01.16 | 7 | 53 | 1 | –1 |
| 48 | Mongrassano (CS) | 39.526 | 16.111 | – | 7/1 | 1908.12.28 | 11 | 152 | 5 | –1 |
| 49 | Montalto Uffugo  (CS) | 39.405 | 16.158 | + | 9/1 | 1990.05.05 | 7 | 174 | 5.5 | +1.5 |
| 50 | Monte Castello di Vibio (PG) | 42.840 | 12.352 | – | 8/2 | 1993.06.05 | 6 | 39 | 1 | –1 |
| | | | | | | 1997.09.26 | 9 | 45 | 4 | –1 |
| 51 | Monteleone,  (VV) Vibo Valentia | 38.675 | 16.102 | – | 10/1 | 1638.03.27 | 11 | 45 | 6.5 | –1 |
| 52 | Montemurro (PZ) | 40.297 | 15.991 | + | 10/2 | 1905.09.08 | 11 | 181 | 6 | +1 |
| | | | | | | 1980.11.23 | 9.5 | 95 | 7 | +1 |
| 53 | Montevarchi (AR) | 43.523 | 11.568 | + | 18/1 | 1915.01.13 | 11 | 246 | 6 | +2 |
| 54 | Narni  (TR) | 42.517 | 12.521 | + | 19/1 | 1998.08.15 | 6 | 47 | 4 | +1 |
| 55 | Nocera Inferiore (SA) | 40.743 | 14.642 | + | 15/1 | 1991.05.26 | 6.5 | 100 | 4.5 | +1 |
| 56 | Nola (NA) | 40.926 | 14.529 | – | 12/2 | 1688.06.05 | 11 | 37 | 5 | –1 |
| | | | | | | 1991.05.26 | 6.5 | 117 | 1 | –1.5 |
| 57 | Notaresco  (TE) | 42.657 | 13.894 | + | 9/1 | 1984.05.07 | 8 | 112 | 6 | +2 |
| 58 | Nusco  (AV) | 40.887 | 15.085 | – | 8/1 | 1930.07.23 | 10 | 25 | 5 | –1 |
| 59 | Osimo  (AN) | 43.485 | 13.483 | + | 22/1 | 1741.04.24 | 9 | 39 | 8 | +1 |
| 60 | Parma (PR) | 44.801 | 10.329 | + | 25/1 | 1989.10.03 | 4 | 35 | 3 | +2 |
| 61 | Paternò  (CT) | 37.566 | 14.902 | + | 13/1 | 1905.09.08 | 11 | 151 | 6 | +1 |
| 62 | Pienza (SI) | 43.079 | 11.679 | + | 8/1 | 1993.06.05 | 6 | 81 | 3 | +2 |
| 63 | Pizzoli (AQ) | 42.435 | 13.303 | + | 11/1 | 1933.09.26 | 9 | 78 | 7 | +1 |
| 64 | Porcia (PN) | 45.964 | 12.618 | + | 10/1 | 1976.05.06 | 9.5 | 46 | 8 | +1 |

| # | Site $g_0$ (province) | coordinates | | type | $f(g)=$ | $I$-map[*)] | | | | |
|---|---|---|---|---|---|---|---|---|---|---|
| | | Lat | Long | $s(g_0)$ | $N_{map}/N_a$ | EQ date | $I_0$ | dist | $I(g_0)$ | $\Delta I(g_0)$ |
| 65 | Porretta (BO) | 44.156 | 10.976 | – | 12/2 | 1909.01.13 | 6.5 | 74 | 3 | –1 |
| | | | | | | 1920.09.07 | 9.5 | 60 | 4 | –1 |
| 66 | Portomaggiore (FE) | 44.697 | 11.805 | + | 13/1 | 1916.05.17 | 7.5 | 105 | 6 | +1 |
| 67 | Prato (PO) | 43.879 | 11.096 | + | 22/1 | 1983.11.09 | 6.5 | 123 | 5 | +1 |
| 68 | Quero (BL) | 45.921 | 11.931 | – | 7/1 | 1895.04.14 | 8.5 | 169 | 4 | –1 |
| 69 | Raccuja (ME) | 38.055 | 14.901 | + | 7/1 | 1905.09.08 | 11 | 115 | 7 | +1 |
| 70 | Rapolla (PZ) | 40.976 | 15.675 | + | 7/2 | 1990.05.05 | 7 | 37 | 7 | +1 |
| | | | | | | 1991.05.26 | 6.5 | 41 | 6.5 | +1 |
| 71 | Rignano Garganico (FG) | 41.675 | 15.587 | – | 8/1 | 1962.08.21 | 9 | 69 | 4 | –2 |
| 72 | Rocca di Papa (RM) | 41.760 | 12.710 | + | 12/2 | 1898.06.27 | 7.5 | 71 | 5 | +1 |
| | | | | | | 1995.06.12 | 5.5 | 24 | 5 | +1 |
| 73 | Rocca San Casciano (FC) | 44.060 | 11.842 | – | 17/1 | 1920.09.07 | 9.5 | 131 | 1 | –2 |
| 74 | Rosarno (RC) | 38.487 | 15.978 | + | 9/1 | 1894.11.16 | 8.5 | 23 | 8 | +1 |
| 75 | Rovigo (RO) | 45.070 | 11.790 | + | 15/1 | 1983.11.09 | 6.5 | 114 | 5 | +1 |
| 76 | San Marco la Catola (FG) | 41.525 | 15.006 | + | 9/1 | 1995.09.30 | 6 | 52 | 5.5 | +1 |
| 77 | San Martino Sannita (BN) | 41.066 | 14.837 | + | 7/2 | 1805.07.26 | 10 | 60 | 8 | +1 |
| | | | | | | 1990.05.05 | 7 | 61 | 6.5 | +1 |
| 78 | San Paolo di Civitate (FG) | 41.739 | 15.261 | – | 8/1 | 1980.11.23 | 9.5 | 101 | 4 | –1 |
| 79 | San Pio delle Camere (AQ) | 42.286 | 13.656 | + | 7/1 | 1990.05.05 | 7 | 242 | 4 | +1 |
| 80 | San Severino Marche (MC) | 43.229 | 13.177 | +/– | 18/4 | 1972.11.26 | 7.5 | 44 | 4 | –1 |
| | | | | | | 1980.11.23 | 9.5 | 341 | 5 | +1 |
| | | | | | | 1987.07.03 | 7 | 40 | 6 | +1 |
| | | | | | | 1993.06.05 | 6 | 39 | 2 | –1 |
| 81 | Sassoferrato (AN) | 43.434 | 12.858 | – | 22/1 | 1984.04.29 | 7 | 32 | 4 | –1 |
| 82 | Sellano (PG) | 42.888 | 12.926 | + | 20/1 | 1987.07.03 | 7 | 59 | 5 | +1 |
| 83 | Staffolo (AN) | 43.432 | 13.186 | +/– | 12/2 | 1741.04.24 | 9 | 18 | 5.5 | –1 |
| | | | | | | 1979.09.19 | 8 | 78 | 6 | +1 |
| 84 | Stornarella (FG) | 41.256 | 15.731 | – | 8/2 | 1913.10.04 | 7.5 | 96 | 1 | –1.5 |
| | | | | | | 1991.05.26 | 6.5 | 71 | 3 | –1 |
| 85 | Subiaco (RM) | 41.925 | 13.095 | + | 19/1 | 1979.09.19 | 8 | 91 | 6 | +1 |
| 86 | Sulmona (AQ) | 42.047 | 13.928 | + | 23/2 | 1933.09.26 | 9 | 18 | 8 | +1 |
| | | | | | | 1987.07.03 | 7 | 127 | 6 | +2 |
| 87 | Toro (CB) | 41.570 | 14.766 | + | 7/1 | 1990.05.05 | 7 | 114 | 4.5 | +1 |
| 88 | Torrecuso (BN) | 41.189 | 14.679 | +/– | 10/2 | 1991.05.26 | 6.5 | 121 | 5 | +2 |
| | | | | | | 1990.05.05 | 7 | 82 | 1 | –3 |
| 89 | Trivigno (PZ) | 40.580 | 15.990 | + | 10/1 | 1905.09.08 | 11 | 213 | 6 | +1 |
| 90 | Urbino (PU) | 43.726 | 12.636 | + | 35/4 | 1873.03.12 | 7.5 | 78 | 6.5 | +1 |
| | | | | | | 1904.11.17 | 7 | 122 | 3 | +2 |
| | | | | | | 1907.01.23 | 5.5 | 154 | 3.5 | +2 |
| | | | | | | 1915.01.13 | 11 | 212 | 6 | +1 |
| 91 | Vacone (RI) | 42.384 | 12.644 | – | 8/1 | 1979.09.19 | 8 | 47 | 4 | –1 |
| 92 | Vallata (AV) | 41.034 | 15.253 | + | 8/1 | 1991.05.26 | 6.5 | 69 | 5.5 | +1 |
| 93 | Vallombrosa (FI) | 43.731 | 11.588 | + | 10/1 | 1914.10.27 | 7 | 94 | 6 | +1 |
| 94 | Velletri (RM) | 41.688 | 12.778 | + | 13/1 | 1990.05.05 | 7 | 255 | 3.5 | +1.5 |
| 95 | Venafro (IS) | 41.485 | 14.044 | + | 15/2 | 1997.03.19 | 6 | 65 | 3 | +2 |
| | | | | | | 1990.05.05 | 7 | 149 | 6.5 | +1.5 |
| 96 | Vicenza (VI) | 45.549 | 11.549 | + | 14/1 | 1983.11.09 | 6.5 | 119 | 5 | +1 |

[*)] EQ, earthquake; $I_0$, intensity at epicenter; dist, distance from $g_0$ in km.